\documentstyle[prl,multicol,aps,psfig]{revtex} 
\begin{document}
\title{Two-electron correlated motion due to Coulomb repulsion}
\author{S.A. Gurvitz}
\address{Weizmann Institute of Science, Department of Particle  
Physics\\ 76100 Rehovot, Israel}
\vspace{18pt} 
\maketitle 
\begin{abstract}
A Hubbard-type model is derived from the microscopic
Schr\"odinger equation. We found that additional terms describing
direct two-electron transitions must be added to the standard
Hubbard Hamiltonian. Such a Hamiltonian
generates two-electron pairing due to on-site 
Coulomb repulsion. We demonstrate that the electron pairs with
opposite spin propagate across periodic structures via direct
and sequential two-electron tunneling. This mechanism can be used
for a generation of entangled electron current. 
Numerical calculations show stability of the electron pairs. 
\end{abstract}
\vspace{10pt} 
\hspace{1.5 cm}   
PACS: 73.63.K,  03.65.X, 71.10.F
\begin{multicols}{2}
Within the framework of the Hubbard model it has been   
earlier shown that the on-side electron-electron interaction ($U$)
might generate coherent propagation of the two electrons\cite{shep,imry}. 
However, in the Hubbard model such a process takes place only via sequential
tunneling, i.e. by a virtual disintegration of a pair. As a
result, the corresponding amplitude is suppressed with increasing $U$.

In fact, a more probable mechanism for two-particle tunneling 
transitions takes place due to simultaneous tunneling of two particles. Yet,  
this type of process is not accounted for by the standard Hubbard Hamiltonian.
In this Letter we show
how such a direct two-electron transitions can be
included in the tunneling Hamiltonian by deriving the latter 
from the original
Schr\"odinger equation. We also demonstrate that such a modified Hubbard
Hamiltonian produces an effective two-electron coupling  
due to on-site electron-electron repulsion. In this case pairs of electron
with opposite spins can propagate along periodic systems, producing a current
of entangled electrons.

We start with an example of two electrons occupying a coupled-dot system,
as shown schematically in Fig.~1.
Each of the dots is represented by a square well potential,
$V_{1,2}(x)$, which contain a bound state at the same energy $E_0$.
If both electrons occupy one of the dots (Fig.~1a,c) the total energy
of this system is  $2E_0+U$. We assume that the inter-dot Coulomb
repulsion is zero, so whenever the electrons are in different dots (Fig.~1b)
the total energy becomes $2E_0$. 

Such a system is usually described by the Hubbard tunneling
Hamiltonian
\begin{equation}  
H = \sum_{i,s}\big (E_0n_{is}+{U\over 2}n_{is}n_{i\bar s}\big )-
\sum_s\Omega_0(a^{\dagger}_{1s}a_{2s}+{\mbox{H.c.}}\big )  
\label{a1}  
\end{equation}
where $a_{is}^{\dagger}$ ($a_{is}$) creates (annihilates) an electron in the 
corresponding dot ($i=1,2$) with the spin $s$ ($\bar s=-s$) and $n_{is}=
a_{is}^{\dagger}a_{is}$. The amplitude $\Omega_0$ generates
transitions between the dots  
via single electron tunneling. This amplitude can
be evaluated by using the Bardeen's formula\cite{bardeen,g1} as a 
product of two bound state wave functions at a point $x_0$ inside
the barrier ($a< x_0<d-a$):
\begin{equation}
\Omega_0=(\kappa /m)\Phi_1(x_0)\Phi_2(x_0)\, .
\label{aa1}
\end{equation}
Here $\Phi_{1,2}(x)$ are the single electron wave functions in the left and
the right dot: $(K+V_{1,2})\Phi_{1,2}=E_0\Phi_{1,2}$ with $K$ is the kinetic
energy operator, and $\kappa =(2m|E_0|)^{1/2}$.
Note that the value of $\Omega_0$ is weakly dependent on $x_0$\cite{g1}.  
For instance, for square well potentials (Fig.~1) and
$a\to 0$, the wave functions 
$\Phi_1(x)=\kappa^{1/2}\exp \left (-\kappa |x|\right )$ and 
$\Phi_2(x)=\kappa^{1/2}\exp \left (-\kappa |d-x|\right )$, so that  
$\Omega_0=(\kappa^2/m)\exp (-\kappa d)$. 
\begin{figure} 
\psfig{figure=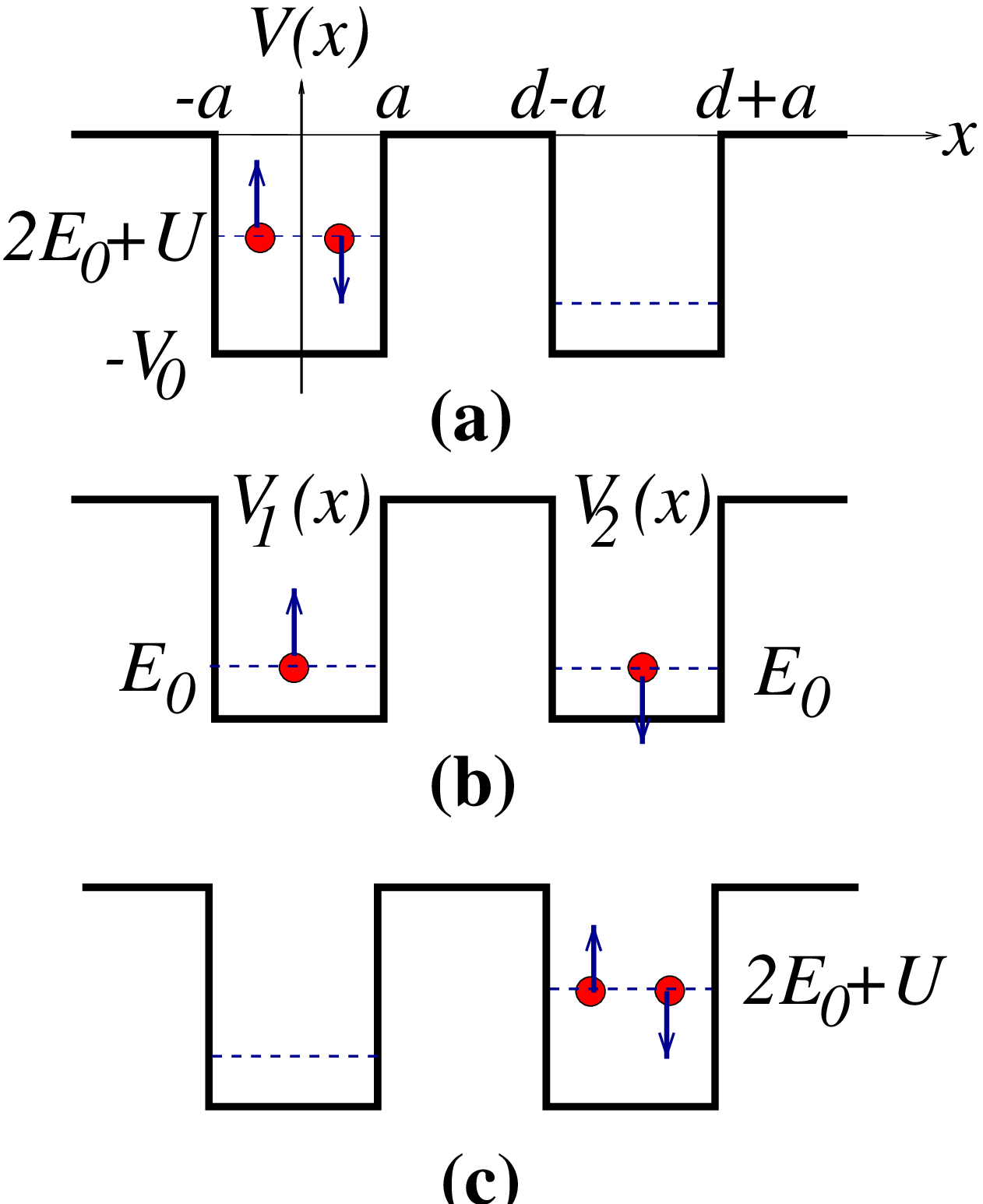,height=6cm,width=6cm,angle=0}
\noindent
{\bf Fig.~1:}
Two electrons with opposite spins in a double-well 
potential $V(x)=V_1(x)+V_2(x)$.
\end{figure}

The wave function $|\Psi (t)\rangle$ describing 
a motion of two electrons inside the
double-dot can be represented by
\begin{equation}  
|\Psi (t)\rangle =\sum_{i,i'=1,2}
b_{ii'}(t)a^{\dagger}_{i\uparrow}
a^{\dagger}_{i'\downarrow}|0\rangle\, , 
\label{a2}  
\end{equation}
where $b_{11}(t), b_{22}(t)$ are the probability amplitudes of
finding two electrons in the first or in the second dot respectively 
(Fig.~1a,c). The probability amplitude $b_{12}(t)$ corresponds to 
the configuration shown in Fig.~1b, and  $b_{21}(t)$ is the same but with 
an interchange of the spins. Substituting Eq.~(\ref{a2}) into the
Schr\"odinger equation $i\partial_t|\Psi (t)\rangle =H|\Psi (t)\rangle$,
we obtain the following coupled equations for the amplitudes $b(t)$
\begin{mathletters}
\label{a3} 
\begin{eqnarray} 
i\dot b_{11(22)}&=&(2E_0+U)b_{11(22)}+\Omega_0(b_{12}+b_{12})
\label{a3a}\\ 
i\dot b_{12(21)}&=&2E_0b_{12(21)}+\Omega_0(b_{11}+b_{22})
\label{a3b}
\end{eqnarray} 
\end{mathletters}

Let us solve these equations for the initial conditions $b_{11}(0)=1$, 
$b_{12(21)}(0)=b_{22}(0)=0$,  
Fig.~1a. Consider first the case of no
Coulomb repulsion between the electrons, $U=0$. Then  
the probability of finding both electrons 
in the same dot, $P_{ii}(t)=|b_{ii}(t)|^2$ 
(Fig.~1a,c), is $P_{11}(t)=\cos^4(\Omega_0t)$ and   
$P_{22}(t)=\sin^4(\Omega_0t)$. Respectively, the probability
of finding two electrons in different dots (Fig.~1b) 
$P_{12}(t)=|b_{12}(t)|^2+|b_{21}(t)|^2$
is $P_{12}(t)=\sin^2(2\Omega_0t)/2$.

However, in the case of strong Coulomb repulsion, $U\gg 4\Omega_0$,   
the time-behavior of two-electron system is very different.
The solutions of Eqs.~(\ref{a3}) 
depend on a small parameter $\Omega_0/U$. 
To second order of this parameter we find 
$P_{11}(t)=\cos^2(\omega t)$, $P_{22}(t)=\sin^2(\omega t)$ with
$\omega =2\Omega^2_0/U\ll\Omega_0$ and  
$P_{12}(t)=(2\Omega^2_0/U^2)\sin^2(Ut/2)\ll 1$.
Thus two electrons oscillate together between
two wells, but with much lower frequency than in the
noninteracting case. The probability of finding two electrons
in different wells (Fig.~1b) is strongly suppressed.   

The above result can be understood in the following way. If a dot is
occupied  by two electrons, the tunneling of one of these electrons out
takes place with a variation of the total energy, 
$\Delta E=-U$, Fig.~1. The probability of such a process is
$\sim \Omega_0^2/\Delta E^2$, and therefore it is strongly suppressed.
This is analogous to the time-evolution of a single particle
in a double-well potential.
The particle remains in the same well where it was initially localized
if the energy levels in the wells are misaligned.  
However if the levels are of the same energy, the particle
oscillates between two wells (Rabi oscillations). In our case
these are the two-electrons states, Fig.~1a,c, which have 
the same energy ($2E_0+U$). Therefore 
the two electrons oscillate between these states.
Yet in the framework of the tunneling Hamiltonian Eq.~(\ref{a1}) these 
oscillations are generated by a sequential tunneling (co-tunneling)\cite{gw}
via the intermediate state of Fig.~1b, so that 
the frequency of oscillations is suppressed by a factor $\Omega_0/U$. 

It follows from these arguments that the probability of 
two-electron tunneling between the states of Fig.~1a,c   
would be enhanced if both electrons tunnel simultaneously,   
keeping the energy of the system constant.
These processes must be accounted
in any tunneling Hamiltonian description
by introducing the corresponding terms generating direct two-particle transitions.
These terms are naturally arise whenever the Hubbard Hamiltonian 
is consistently derived from the Schr\"odinger equation.  
Such a derivation is presented below by using the Green's
function technique of the two-potential approach\cite{g1,g2}.   

Consider the Schr\"odinger equation for two
electrons in  a double-well potential $V(x)=V_1(x)+V_2(x)$, Fig.~1.
For simplicity we
disregard the spin by treating two electrons with opposite spin as
distinguishable particles. The total Hamiltonian can be written as
$H=H_1+H_2+U_C$ with  $H_i=K(x_i)+V(x_i)$ for $i=1,2$,
and $U_C(x_1-x_2)$ is the Coulomb repulsive potential. 
We assume that $U_C=U$ for $|r_1-r_2|\lesssim \bar c$
and drops down outside this region, where $a\ll  \bar c\lesssim d$. 

We start with two electrons localized in the first dot,  
$|\Psi (0)\rangle =|\Phi_{11}\rangle$, where $\langle x_1,x_2|\Phi_{11}\rangle
=\Phi_1(x_1)\Phi_1(x_2)$. To determine the
time-development of this system we apply the Laplace transform
$|\tilde\Psi (E)\rangle =\int_0^{\infty}|\Psi (t)\rangle\exp (iEt)dt$.
The Schr\"odinger equation then reads
\begin{equation}
(E-H)|\tilde\Psi (E)\rangle =i|\Psi (0)\rangle 
\label{aa3}
\end{equation}
Using the Green's function technique\cite{g1,g2}, we can rewrite the 
total wave function as 
\begin{eqnarray}
|\tilde\Psi (E)\rangle =[1+\tilde G(E)\bar V_2]
\tilde b_{11}(E)|\Phi_{11}\rangle\,  
\label{a4}\\[5pt]
\tilde G(E)=G_1(E)(1-\Lambda_{11})[1+\bar V_2\tilde G(E)]\, ,
\label{a5}
\end{eqnarray}
where $G_1(E)=(E-\bar K-\bar V_1)^{-1}$ with 
$\bar K=K(x_1)+K(x_2)$, $\bar
V_{1,2}=V_{1,2}(x_1)+V_{1,2}(x_2)+U_C^{(1,2)}(x_1-x_2)$ and
$\Lambda_{11}=|\Phi_{11}\rangle\langle\Phi_{11}|$ is the projection operator.
Here we represented $U_C= U_C^{(1)}+U_C^{(2)}$, where $U_C^{(1)}=U_C$
for $x_1,x_2\lesssim \bar c$ and drops to zero outside this region.
Correspondingly, $U_C^{(2)}=U_C$
for $d-x_1,d-x_2\lesssim \bar c$ and drops to zero outside this region.
The amplitude $\tilde b_{11}(E)$ in Eq.~(\ref{a4}) 
is the Laplace transform of the probability amplitude $b_{11}(t)$ 
of finding both electrons in the first dot, Fig.~1a. One obtains from
Eqs.~(\ref{aa3})-(\ref{a5})\cite{g1} 
\begin{equation} \tilde b_{11}(E)={i\over
E-2E_0-U-\langle\Phi_{11}| \bar V_2+\bar V_2\tilde G(E)V_2|\Phi_{11}\rangle} 
\label{aa5}
\end{equation}

Eq.~(\ref{a4}) can be treated iteratively by expanding  $\tilde G$,
Eq.~(\ref{a5}), in powers of $G_1$. Since $E\simeq 2E_0+U$
the Green's function $G_1$ is dominated by the bound
state pole in its spectral representation,  
$G_1(E)\to |\Phi_{11}\rangle\langle\Phi_{11}|/(E-2E_0-U)$ 
for $E\to 2E_0+U$. Yet, this pole is excluded by the projection operator 
$1-\Lambda_{11}$. The remaining part of $G_1$ gives rise to 
the corresponding (Born) series for $|\tilde\Psi (E)\rangle$ which, however,
converges very slowly.  We therefore look
for a different expansion for $\tilde G$ which converges much faster
than the Born series. This can be achieved by expanding  $\tilde G$ in powers
of  the Green's function $G_2(E)=1/(E-\bar K-\bar V_2)$
by using the relation\cite{g1,g2}
\begin{equation}
\tilde G=G_2(1+\bar V_1\tilde G)
-G_2\Lambda_{11}(1+\bar V_2\tilde G) \, . 
\label{a6}
\end{equation}
Since the second dot contains the two-electron bound state at the
same energy $2E_0+U$ as the first dot, the Green's function 
$G_2(E)$ can be replaced by 
$G_2(E)\to |\Phi_{22}\rangle\langle\Phi_{22}|/(E-2E_0-U)$ 
for $E\to 2E_0+U$, where $\langle x_1,x_2|\Phi_{22}\rangle
=\Phi_2(x_1)\Phi_2(x_2)$. Substituting this expression into
Eq.~(\ref{a6}) we find that in the contrast to Eq.~(\ref{a5})
the bound state pole is not cancelled by the projection operator $1-\Lambda_{11}$. 
In addition, the second term in Eq.~(\ref{a6}) 
is proportional to the overlap $\langle\Phi_{22}|\Phi_{11}\rangle$ and
can be considered as a small perturbation. If we neglect this term
Eq.~(\ref{a6}) can be easily solved thus obtaining
\begin{equation}
\tilde G(E)\simeq {|\Phi_{22}\rangle\langle\Phi_{22}|\over E-2E_0-U-
2\delta_{12}}\, ,
\label{a7}
\end{equation}
where $\delta_{12}=\langle\Phi_2|V_1|\Phi_2\rangle$ is a small energy shift. 
Note that the Green's function $\tilde G(E)$ contains also the poles
at $E\simeq 2E_0$, corresponding to separated electrons, 
Fig.~1b. Since $E\simeq 2E_0+U$ these poles do not affect
$\tilde G(E)$ for large $U$. 
Substituting Eq.~(\ref{a7}) into Eq.~(\ref{a4}) we find
\begin{equation}
|\tilde\Psi (E)\rangle = \tilde b_{11}(E)|\Phi_{11}\rangle
+\tilde b_{22}(E)|\Phi_{22}\rangle\, ,
\label{a8}
\end{equation}
where
\begin{equation}
\tilde b_{22}(E)  
=-{\Omega_2\over E-2E_0-U-
2\delta_{12}}\tilde b_{11}(E)  
\label{a9}
\end{equation}
is the Laplace transform of the probability amplitude
$b_{22}(t)$, Eq.~(\ref{a2}), and 
$\Omega_2=-\langle\Phi_{22}|\bar V_2|\Phi_{11}\rangle$.
Then substituting Eq.~(\ref{a7}) into Eq.~(\ref{aa5})
and using Eq.~(\ref{a9}) we obtain 
\begin{equation}
(E-2E_0-U-2\delta_{21})\tilde b_{11}(E)
+\Omega_2\tilde b_{22}(E)=i\, , 
\label{a10}
\end{equation}
where $\delta_{21}=\langle\Phi_1|V_2|\Phi_1\rangle$
(in our case $\delta_{12}=\delta_{21}=\delta$).

Performing the inverse Laplace transform,  
Eqs.~(\ref{a9})-(\ref{a10}) can be rewritten in a matrix form as
\begin{equation}
i\left(\begin{array}{l}
\dot b_{11}(t) \\\dot b_{22}(t)
\end{array}\right)=
\left(\begin{array}{cc}
2E'_0+U & -\Omega_2 \\
-\Omega_2 & 2E'_0+U
\end{array}
\right )\left(\begin{array}{l}
b_{11}(t) \\b_{22}(t)
\end{array}\right)\, ,
\label{a11}
\end{equation}
where $E'_0=E_0+\delta$. 
Using the occupation number representation and re-introducing
the spin variable one finds that Eq.~(\ref{a11}) can be rewritten as the 
Schr\"odinger equation $i\partial_t|\Psi (t)\rangle =H|\Psi
(t)\rangle$, where $|\Psi (t)\rangle$ is given by Eq.~(\ref{a2})
and the Hamiltonian 
\begin{eqnarray}  
H = \sum_{i=1,2}\big [E'_0(a^{\dagger}_{i\uparrow}a_{i\uparrow}&+&
a^{\dagger}_{i\downarrow }a_{i\downarrow})+
Ua^{\dagger}_{i\uparrow}a_{i\uparrow}a^{\dagger}_{i\downarrow}
a_{i\downarrow}\big ]
\nonumber\\
&-&~\Omega_2(a^{\dagger}_{1\uparrow}a^{\dagger}_{1\downarrow}
a_{2\uparrow}a_{2\downarrow}
+H.c.)\, .
\label{a12}  
\end{eqnarray}
Here the amplitude $\Omega_2=-\langle\Phi_{22}|\bar V_2|\Phi_{11}\rangle$
generates a direct coupling in 2-dimensional Hilbert space
between two-electron states, Fig.~(1a)$\leftrightarrow$Fig.~(1c). This coupling can be
calculated directly\cite{imry1} by
using $\langle\Phi_{22}|\bar V_2
=\langle\Phi_{22}|(2E_0-\bar K )$ and integrating by parts. As a
result we find the following simple expression (c.f.\cite{bardeen,g1,g2}).  
\begin{equation}
\Omega_2={2\kappa\over m}\Phi_1(x_0)\Phi_2(x_0)
\int\Phi_1(x')\Phi_2(x')dx'\, .  
\label{a13}
\end{equation}
For the square-well potentials (Fig.~1) 
and $a\to 0$, Eq.~(\ref{a13}) yields  
$\Omega_2=(\kappa d/|E_0|)\Omega_0^2$, where $\Omega_0$ 
is single-electron tunneling amplitude, Eq.~(\ref{aa1}).
As expected the direct two-electron transitions, generated by  
$\Omega_2$, dominate over the sequential two-electron  
transitions, $\sim 2\Omega_0^2/U$, with an increase of $U$.
This dominance of course is not limited to square-well potentials. 

Note that Eq.~(\ref{a13}) has been derived for long range repulsive
potential $U_C$. For short range potential such that 
$U_C(x_1-x_2)\to 0$ for $|x_1-x_2|\lesssim a$, the wave function
$\Phi_{11}$ and $\Phi_{22}$ are not given by a product
of single electron wave functions. Yet the amplitude
$\Omega_2$ can be evaluated by using the Bardeen formula\cite{bardeen,g2}.

Comparing Eq.~(\ref{a12}) with  Eq.~(\ref{a1}) we find that
the term $\Omega_0a^\dagger_{1s}a_{2s}$
describing single electron transitions does not arise
in Eq.~(\ref{a12}) since the corresponding poles of $\tilde G(E)$
at $E\simeq 2E_0$ were neglected in
our derivation. However with a decrease of $U$, such that
$\Omega_2\sim 2\Omega_0^2/U$, 
these poles must be taken into account. 
As a result we arrive to the Hamiltonian (\ref{a1}) supplemented with the last
term of Eq.~(\ref{a12}). 
Solving the Schr\"odinger equation with such a modified Hamiltonian
for the initial conditions corresponding to Fig.~1a we find 
in the limit of $U\gg\Omega_0$ that 
$P_{11}(t)=\cos^2(\omega't)$, $P_{22}(t)=\sin^2(\omega't)$
and $P_{12}(t)=(2\Omega_0^2/U^2)\sin^2(Ut/2)$, where
$\omega'=|\Omega_2-(2\Omega_0^2/U)|$. Note that the transition
amplitudes $\Omega_0$ and $\Omega_2$, given by Eqs.~(\ref{aa1} )
and (\ref{a13}) change the sign if the wave functions
$\Phi_{1,2}(x)$ are of a different parity. 
 
Our procedure can be extended to a general case of $N$ coupled wells.
Detailed derivation will be presented elsewhere. Here we give 
our final result, representing the Hubbard-type Hamiltonian
with additional terms for direct two-electron transitions 
\begin{eqnarray}  
H=\sum_{i,s}^N&&\big ( E'_0n_{is}
+{U\over 2}n_{is}n_{i\bar s}\big )-\sum_{i,s}^{N-1}
\big (\Omega_0a^{\dagger}_{is}a_{i+1,s}
\nonumber\\
&&+{\Omega_2\over 2}a^{\dagger}_{is}a^{\dagger}_{i\bar s}
a_{i+1,s}a_{i+1,\bar s}
+{\mbox{H.c.}}\big )\, .
\label{a15}  
\end{eqnarray}
Here $i=1,\ldots ,N$ and $E'_0=E_0+\delta$, where $\delta$ is a small energy
shift defined in the same way as in Eq.~(\ref{a11}). 
The two-electron coupling $\Omega_2$ is given by
Eq.~(\ref{a13}) for $U\gg\Omega_0$, but $\Omega_2\to 0$ for $U\to 0$.
In this limit, however, the contribution
from direct two-electron transitions is suppressed with respect
to that generated by single electron transitions. Thus  
Eq.~(\ref{a15}) with $\Omega_2$ given by Eq.~(\ref{a13})
can be used for any values of $U$. 

As in the previous case of a double-well potential,
an electron pair with opposite spins initially localized
in one of the wells cannot
be separated in the limit of $U\to\infty$\cite{fn}. Such a pair 
can only move as a whole
object due to direct two-electron tunneling between neighboring
wells (the amplitude of co-tunneling vanishes in this limit).
As a result a mini-band of the width $2\Omega_2$
appears, providing a current of entangled electrons.

If $\Omega_2\sim 2\Omega_0^2/U$, the propagation of an electron pair is govern
by the both co-tunneling and direct two-electron transitions.
In order to assess their importance with respect to the single electron transport 
we solved numerically the Schr\"odinger equation $i\partial_t|\Psi (t))
\rangle =H|\Psi (t)\rangle$ with the Hamiltonian (\ref{a15})
and the wave function given by Eq.~(\ref{a2}) with $i,i'=1,\ldots ,N$  
for $N=10$ and $U=10\Omega_0$ and the initial conditions corresponding
to the first site occupied by two electrons.
The results of our calculations are shown in Fig.~2. We plot there
the probabilities of finding one and two electrons at the last site,
$P_{N}(t)$ and $P_{NN}(t)$, for $\Omega_2=0.1\Omega_0$ and $\Omega_2=0$. 
Although for the chosen values of parameters
the value of $\Omega_2=0.1\Omega_0$ is smaller than that of the co-tunneling
one ($2\Omega_0^2/U$), the direct two-electron transitions strongly affects
the pair transport. For instance the pair reaches the last site
considerably faster and with much larger probability.
\begin{figure} 
\psfig{figure=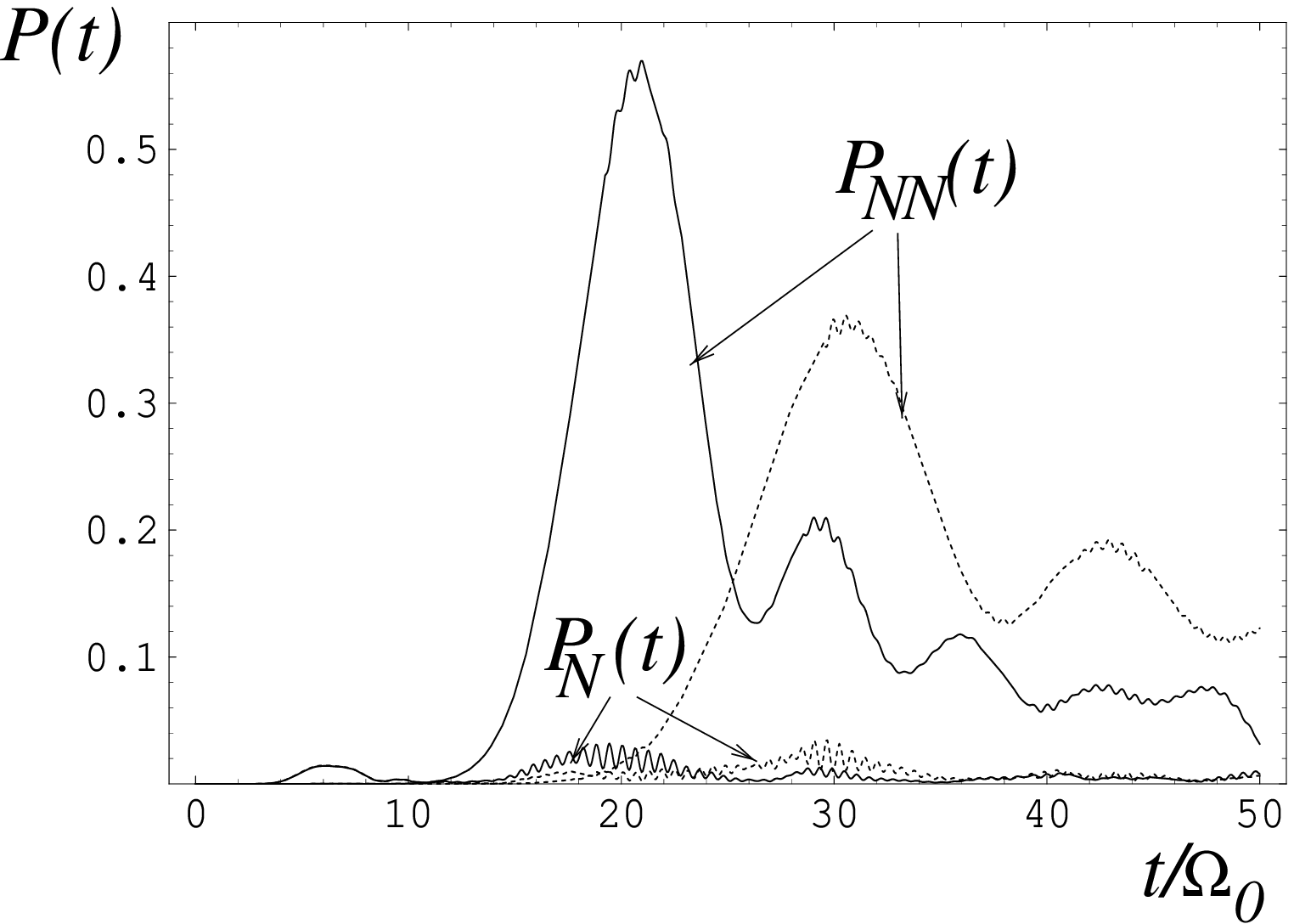,height=6cm,width=8cm,angle=0}
\noindent
{\bf Fig.~2:}
The probability of finding one and two electrons at the
last site as a function of time. The solid lines correspond
to $\Omega_2=0.1\Omega_0$ and the dotted lines to $\Omega_2=0$. 
\end{figure}

It follows from Fig.~2 that  
the probability of finding a single electron at the last site 
is much lower than that of an entangled electron pair. 
Yet, one cannot establish from this figure that
this is true for all sites. Moreover, it is 
natural to assume that the electron pair decays very fast.
Indeed, the weight of all states with two electrons are 
at different sites is much larger then those occupying
the same site. In order to investigate this point we evaluated numerically 
the total probability of finding two electrons at the same site inside
the chain, $P_{pair}(t)=\sum_i|b_{ii}(t)|^2$.
The results for the previous values of parameters
are displayed in Fig.~3. It follows from this figure that
contrary to the expectations the two-electron pair is 
stable during long time interval. It is interesting
to note that the direct two-electron transitions affect the 
stability of this pair very little. 

The entangled electron pair transport discussed above 
can be realized in multi-dot structures  
coupled to the reservoirs (emitter and collector)
with a voltage bias larger than $U$.
Then two electrons with opposite spins can enter 
simultaneously the first  
dot and then propagate across the entire structure.
The single and the pair electron currents 
in the collector can be found in the most simple way by 
using the Bloch-type rate equations for
the reduced density matrix derived in\cite{gp}.
\begin{figure} 
\psfig{figure=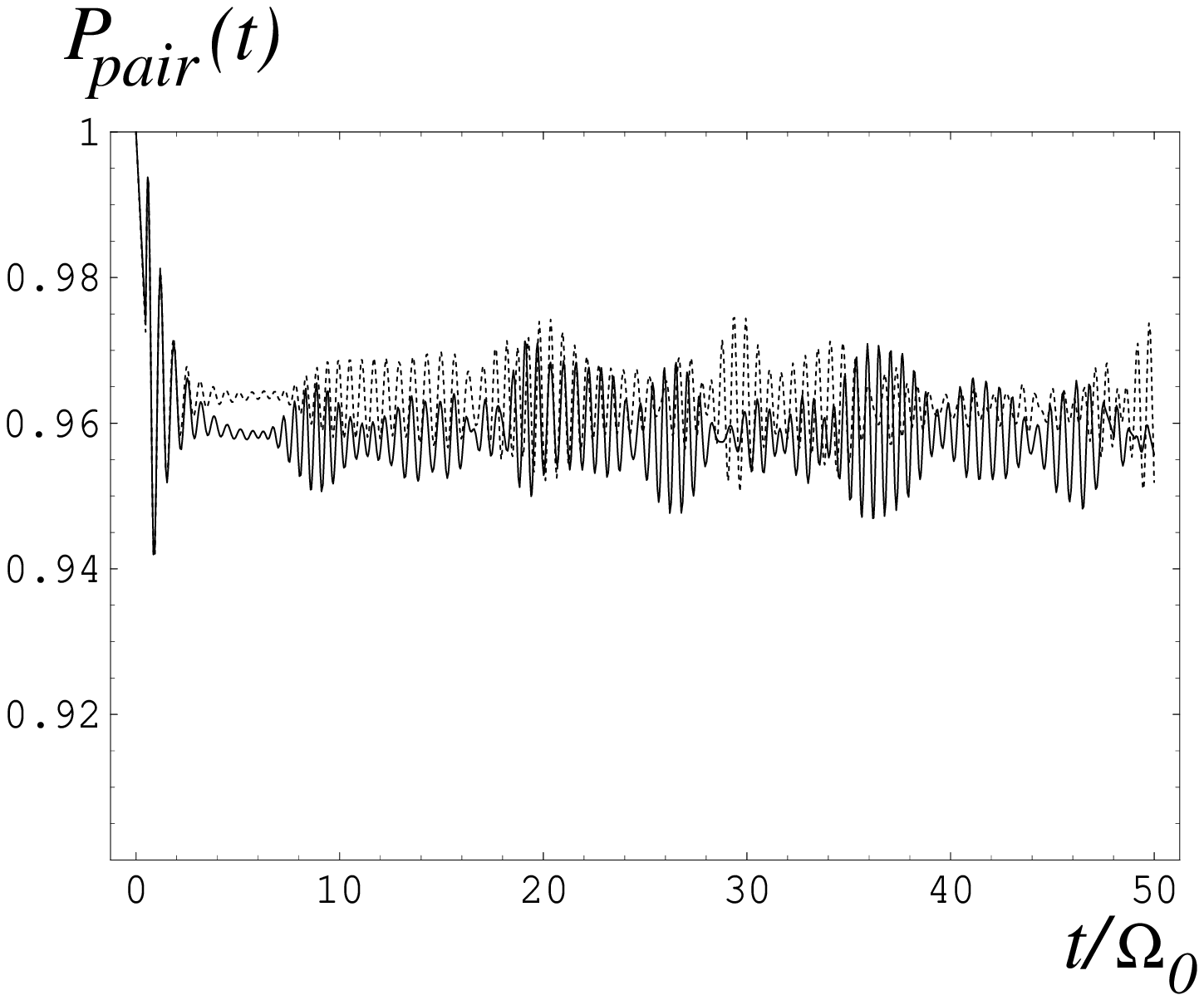,height=6cm,width=8cm,angle=0}
\noindent
{\bf Fig.~3:}
Total probability of finding two electrons bound 
as a function of time, $P_{pair}(t)$.
The solid and dotted lines correspond to $\Omega_2=0.1\Omega_0$
and $\Omega_2=0$, respectively.
\end{figure}

The two-electron coupling can also take place between electrons in
neighboring dots via the inter-dot Coulomb repulsion.
Indeed, if the separation energy of such a pair, $U_{i,i+1}-U_{i,i+2}$
is larger then a single-electron transition amplitude, $\Omega_0$,
the pair moves across the system as a whole object. This phenomenon 
for three coupled dots has been studied in\cite{gw}, but 
only taking into account the co-tunneling mechanism for the pair
transport.

In conclusion, we demonstrated that the Coulomb repulsion
can bound electrons in multi-well periodic structures,
so that a current of entangled electrons with opposite
spins would appear. The propagation
of such electron pairs across the system
is dominated by direct two-electron transitions absent
in the standard Hubbard model. We expect
that these two-electron transitions play an important role in
different physical processes.

The author is indebted to M. Heiblum, Y. Imry, M. Kugler and
A. Yacoby for fruitful discussions.

\end{multicols} 
\end{document}